\documentclass[%
 aip,
 amsmath,amssymb,
 reprint,%
]{revtex4-1}

\usepackage{graphicx}
\usepackage{dcolumn}
\usepackage{bm}

\usepackage[utf8]{inputenc}
\usepackage[T1]{fontenc}
\usepackage{lineno}
\usepackage{mathptmx}
\usepackage{etoolbox}
\usepackage{amsmath}
\usepackage{braket}
\usepackage{xcolor}

\makeatletter
\def\@email#1#2{%
 \endgroup
 \patchcmd{\titleblock@produce}
  {\frontmatter@RRAPformat}
  {\frontmatter@RRAPformat{\produce@RRAP{*#1\href{mailto:#2}{#2}}}\frontmatter@RRAPformat}
  {}{}
}%
\makeatother
\begin{document}

\newcommand{\xhat}{{\bf \hat{x}}}
\newcommand{\yhat}{{\bf \hat{y}}}
\newcommand{\zhat}{{\bf \hat{z}}}
\newcommand{\ehat}{{\bf \hat{e}}}
\newcommand{\BE}{{\bf E}}

\preprint{AIP/123-QED}

\title {Simple experimental realization of optical Hilbert Hotel using scalar and vector fractional vortex beams}
\author{Subith Kumar}
\affiliation{Physical Research Laboratory, Ahmedabad, Gujarat  380009, India}
\affiliation{Indian Institute of Technology Gandhinagar, Ahmedabad, Gujarat 382424, India}
%
\author{Anirban Ghosh}%
\altaffiliation[Author to whom correspondence should be addressed.]{}
\affiliation{Physical Research Laboratory, Ahmedabad, Gujarat  380009, India}
\email{anirbang@prl.res.in, gjgbur@uncc.edu, gsamanta@prl.res.in}
\author{Chahat Kaushik}
\affiliation{Physical Research Laboratory, Ahmedabad, Gujarat  380009, India}
\affiliation{Indian Institute of Technology Gandhinagar, Ahmedabad, Gujarat 382424, India}
\author{Arash Shiri}
\affiliation{Department of Physics and Optical Science, University of North Carolina Charlotte, Charlotte, North Carolina 28223}
\author{Greg Gbur}
\affiliation{Department of Physics and Optical Science, University of North Carolina Charlotte, Charlotte, North Carolina 28223}
%
\author{Sudhir Sharma}
\affiliation{Dynotech Instruments Pvt Ltd, Delhi 110058, India}
\author{G. K. Samanta}
\affiliation{Physical Research Laboratory, Ahmedabad, Gujarat  380009, India}

\date{\today}

\begin{abstract}
Historically, infinity was long considered a vague concept -- boundless, endless, larger than the largest -- without any quantifiable mathematical foundation. This view changed in the 1800s through the pioneering work of Georg Cantor showing that infinite sets follow their own seemingly paradoxical mathematical rules.  In 1924, David Hilbert highlighted the strangeness of infinity through a thought experiment now referred to as the Hilbert Hotel paradox, or simply Hilbert's Hotel. The paradox describes an ``fully” occupied imaginary hotel having infinite number of single-occupancy rooms, the manager can always find a room for new guest by simply shifting current guests to the next highest room, leaving first room vacant.  
The investigation of wavefield singularities has uncovered the existence of a direct optical analogy to Hilbert's thought experiment. Since then, efforts have been made to investigate the properties of Hilbert's Hotel by controlling the dynamics of phase singularities in ``fractional'' order optical vortex beams. Here, we have taken such proposals to the next level and experimentally demonstrated Hilbert's Hotel using both phase and polarization singularities of optical fields. Using a multi-ramped spiral-phase-plate and a supercontinuum source, we generated and controlled fractional order vortex beams for the practical implementation of Hilbert's Hotel in scalar and vector vortex beams. Using a multi-ramped spiral-phase-plate, we show the possibility for complicated transitions of the generalized Hilbert's Hotel. The generic experimental scheme illustrates the usefulness of structured beams in visualizing unusual mathematical concepts and also for fractional vector beams driven fundamental and applied research.

\end{abstract}

\maketitle

\section{Introduction}

The study of wavefield singularities in light, now referred to as singular optics \cite{mssmvv:pio}, has revealed many interesting phenomena and new applications. In scalar waves, these singularities typically manifest as lines of zero intensity in three-dimensional space around which the phase of the field has a circulating or helical structure, as first demonstrated by Nye and Berry \cite{jfnmvb:pra:1974}; such structures are now referred to as optical vortices. In a closed path around an optical vortex, the phase always changes by an integer multiple of $2\pi$; this multiple is referred to as the topological charge. In vector waves, singularities manifest as lines of circular polarization in three-dimensional space, upon which the orientation of the major axis of the polarization ellipse is undefined; the typical form of these singularities are usually referred to as C-lines \cite{jfn:prsla:1983}.  In a closed path around a C-line, the orientation changes by a half-integer multiple of $2\pi$, and this multiple is called the topological index.  Phase and polarization singularities typically intersect a transverse plane of an optical beam at a point; for polarization, we then refer to C-points.

Optical singularities have been considered for a number of applications due to their unusual properties. Beams carrying a pure optical vortex on their central axis, such as Laguerre-Gauss beams, possess a well-defined orbital angular momentum (OAM), and this OAM has been used for the trapping and rotation of particles and creation of light-driven micromachines \cite{kldgg:oe:2004}.  Pure OAM states can be multiplexed and demultiplexed, and there are numerous investigations in using an OAM basis to increase the data transmission rate in optical communications \cite{ggjjcmjpmvvpsmbsfa:oe:2004,jwyyyimfnayyhhyryysdmtaew:np:2012}. Both the topological charge of vortices and the topological index of C-lines are stable under weak perturbations of the wavefield, and typically only created or destroyed in equal and opposite pairs; because of this, these structures have been studied as alternative information carriers in optical communications \cite{ggrkt:josaa:2008}.

One particularly surprising discovery to come from investigations of wavefield singularities is the existence of an optical analogy to the mathematics of transfinite numbers. In a famous lecture, mathematician David Hilbert introduced what is now known as Hilbert's Hotel to highlight how strange the mathematics of infinity would be in a real-world setting; his description was later popularized by Gamow \cite{gg:otti}. Hilbert imagined a Hotel with a countably infinite number of rooms, numbered $1,2,3,\ldots$, all occupied, so the Hotel has no vacancies. However, each guest can be asked to move to the next highest room, making room 1 available, and this process can be repeated indefinitely. Hilbert's Hotel, therefore, simultaneously has no vacancies and an infinite number of vacancies.

Despite the paradoxical nature of Hilbert's Hotel, it has now been recognized to manifest in systems that possess wavefield singularities, with singularities of positive and negative topological charge (or index) representing the ``guests'' and ``rooms'' of the Hotel. The first hint of this behaviour was shown by Berry \cite{mvb:joa} in 2004, who noticed that the creation of vortices by a fractional order spiral phase plate goes through a state with an infinite number of vortex pairs when the effective order of the plate is half-integer, with the topological charge of the field discontinuously changing at that moment; his results were confirmed experimentally the same year \cite{jleymjp:njp:2004}. In 2016, Gbur \cite{gg:o:2016} argued that this system violates the conservation of topological charge through a mechanism directly analogous to Hilbert's Hotel and shows that multi-ramp spiral phase plates use the Hotel effect to create multiple vortices simultaneously. Furthermore, in 2017, Wang and Gbur \cite{Wang:17} showed theoretically that novel Hotel effects could be created with polarization singularities in vector beams.  The creation of infinite pairs of singularities in space is not the only way to realize Hilbert's Hotel with OAM states; in 2015, researchers demonstrated a version of Hilbert's Hotel through multiplicative mapping of OAM modes \cite{vpfmmmmosmlacldklorwbjj:prl:2015}.

With such non-intuitive phenomena predicted, and the possibility of discovering more unusual vortex phenomena related to transfinite mathematics, it is worthwhile to have robust and versatile experimental techniques for testing these effects. Recently, the optical vortex version of Hilbert's Hotel for a single ``room'' was demonstrated experimentally using a spatial light modulator (SLM) to produce the fractional vortex states \cite{PhysRevA.106.033521}. The SLM was encoded with the desired fractional order and illuminated by a Gaussian beam to produce a fractional vortex beam. However, detailed investigations of Hilbert's Hotel require nearly continuous changes in the fractional vortex order, and despite the flexibility of SLMs in terms of dynamic phase modulation and wide wavelength coverage, it would be advantageous to have methods of studying the phenomenon that do not rely on their discrete nature. Furthermore, high-power vortex beams used in many applications are typically generated using spiral phase plates (SPPs), refractive elements that use a helical ramp structure to generate a vortex. If the unusual effects of the optical Hilbert's Hotel find practical application, having methods of generating them using SPPs would be beneficial.

SPPs were originally designed to produce a beam with a specific topological charge at a desired wavelength. Based on the design wavelength, $\lambda$, and the target vortex order, $l$, the maximum step height $s$ is engineered in such a way that the incident Gaussian beam acquires an azimuthal shift of $2\pi l$. The vortex order of the beam is related to the optical parameters by the formula,
\begin{equation}
  \label{eq:E1}
 l = [n(\lambda) - 1]s/\lambda,
\end{equation}
where $n(\lambda)$ is the wavelength-dependent refractive index of the material. As a result, SPPs are very wavelength specific and will not generate an integer vortex beam of order $l$ for a laser wavelength away from the designed wavelength, $\lambda$. This limitation turns out to be an advantage for demonstrating Hilbert's Hotel: from Eq.~(\ref{eq:E1}) it is evident that one can continuously vary the effective vortex order of the SPP from $l$ to $2l$ by varying the beam wavelength from $\lambda$ to $\lambda/2$, with a small modification due to the variation of refractive index with wavelength. Therefore, by simply changing the wavelength of a Gaussian beam incident upon an SPP, one can easily generate vortex beams that effectively have a continuously variable fractional charge.

In this paper we introduce, for the first time to the best of our knowledge, a simple experimental technique for studying fractional vortex effects, including not only the original vortex Hilbert's Hotel but also its vector beam generalization. Using a double-ramp SPP having a step height corresponding to the vortex order $l=2$ at the designed wavelength of 1064 nm in a modified Mach-Zehnder interferometer, and a supercontinuum laser tunable across 400-800 nm, we have generated scalar and vector vortex beams with tunable fractional topological order and verified several varieties of the optical Hilbert's Hotel.

\section{Theoretical framework of Hilbert's Hotel Evolution in fractional singularities}

Before describing the experiment, we briefly review the mathematics related to fractional spiral phase plates for both scalar phase singularities and vector polarization singularities.  The results of these calculations are shown in Section \ref{experimentsetup} to compare with experimental measurements.

\subsection{Phase Singularities}

A common method of generating a vortex beam is to illuminate a spiral phase plate with a normally incident plane wave. As noted in the Introduction, a standard SPP has a helical ramp structure that, for a given wavelength, produces a continuous phase change $2\pi \alpha$ in the azimuthal direction; we refer to $\alpha$ as the order of the SPP. Beams produced by a SPP will always have an integer topological charge, regardless of whether $\alpha$ is integer or fractional. For brevity, however, we will refer to $\alpha$ as the order of the transmitted beam.

We first consider a unit amplitude scalar plane wave passing through a SPP with integer order $n$ and transmission function $\exp(in\phi)$. At a distance $z$ from the phase plate, the transmitted field can be determined by Fresnel diffraction, and is of the form \cite{mvb:joa},
\begin{equation}\label{1}
    \begin{split}
        U_n (r , \phi , z)=&\frac{\pi}{2} \sqrt{\frac{(-i)^{|n|}}{\lambda z}} r\exp(in\phi)\exp(ikz) \exp\left(\frac{ikr^2}{4z}\right)\\
        \times& \left[ J_{\frac{|n|-1}{2}}\left(\frac{kr^2}{4z}\right) - iJ_{\frac{|n|+1}{2}}\left(\frac{kr^2}{4z}\right)\right]\, ,
    \end{split}
\end{equation}
where $\mathbf{r}=(r,\phi)$ denotes the position vector in the transverse plane, $\lambda$ and $k$ denote the wavelength and wave number of the beam, respectively, and  $J_n$ represents the $n^{th}$ order Bessel function of the first kind.

Traditional SPPs have a single ramp along the azimuthal direction and a single step discontinuity; we consider the more general case given by Gbur \cite{gg:o:2016} of a SPP with $m$ ramps of equal azimuthal width $\varphi_m=2\pi/m$, and overall fractional order $\alpha$. The transmission function of the SPP is given by
\begin{equation}\label{2}
    T(\phi) = \exp[i\alpha(\phi - p\varphi _m )],   \quad p\varphi_m\leq 
    \phi \leq (p+1)\varphi_m,
\end{equation}
where $p=0, 1, 2,..., m-1$ and $\alpha$ can have any integer or fractional value.

To evaluate the propagation of a plane wave through this fractional SPP, we may expand the transmission function of the multi-ramp fractional SPP in a Fourier series,
\begin{equation}\label{3}
    T(\phi) =  \sum_{n=-\infty}^{\infty} C_n e^{in\phi} , 
\end{equation}
where the expansion coefficients are readily found to be of the form,
\begin{equation}\label{4}
    C_n =  \frac{i}{2\pi}\sum_{p=0}^{m-1} e^{-ipn\varphi_m}\left[ \frac{1-e^{i\varphi_m(\alpha-n)}}{\alpha-n}\right].
\end{equation}
Equation (\ref{3}) expresses the transmission function of the fractional SPP as a superposition of integer SPP transmission functions. The fractional scalar vortex beam $U_\alpha(\mathbf{r},\phi,z)$ generated by the fractional SPP can then also be written as the superposition of integer order vortex beams with different amplitudes (expansion coefficients) depending on the order $\alpha$ of the SPP,
\begin{equation}\label{5}
    U_\alpha(r,\phi,z) = \sum_{n=-\infty}^{\infty} C_n U_n(r,\phi,z) .
\end{equation}

It was shown by Gbur that the net topological charge of the transmitted field increases by integer multiples of $m$, and the transition happens when the fractional order of the SPP is an odd multiple of $m/2$.

In our experiment, we have used a double ramp SPP $(m=2)$ with the fractional order
\begin{equation}\label{6}
   \alpha=\frac{s}{\lambda}\left[n(\lambda)-\frac{m}{2}\right], 
\end{equation}
 where $n(\lambda)$ is the wavelength dependent refractive index of the BK7 glass, determined by the empirical Sellmeier formula \cite{BK7}. The maximum height difference in the SPP is given by the relation 
\begin{equation}\label{6a}
   s=\left[\frac{l_{d} \lambda_{d}}{n(\lambda_{d})-\frac{m}{2}} \right],
\end{equation}
 where $l_d$ and $n(\lambda_d)$ are the vortex order and refractive index of the SPP at the designed wavelength, $\lambda_d$ = 1064 nm. By continuously modulating the wavelength of the source, the optical path length difference the light experiences on transmission will change and hence its output phase can change by fractional values of $2\pi$, implying the generation of fractional vortex beams of varying order.\par
 
 \subsection{Polarization Singularities}
 
 According to Wang and Gbur \cite{Wang:17}, a similar mathematical method can be applied to generating vector beams with an effective fractional topological index. In the experiment described in Section \ref{experimentsetup}, we superimpose a right-hand circularly (RHC) polarized Gaussian beam with a left-hand circularly (LHC) polarized beam that has passed through the fractional SPP. The output field may therefore be written in the form,
 \begin{equation}\label{E:pm}
     \mathbf{E}_\alpha(r,\phi) = U_\alpha(r,\phi)\ehat_++U_0(r,\phi)\ehat_-,
 \end{equation}
 where $\ehat_+$ and $\ehat_-$ are the left- and right-handed circular polarization unit vectors, 
 \begin{equation}
     \ehat_\pm = \xhat\pm i \yhat.
 \end{equation}
In Eq.~(\ref{E:pm}), C-points can be readily identified as points where $U_\alpha(r,\phi)$ = 0, resulting in a point of pure circular polarization.

In terms of linear polarization, we may write
 \begin{equation}
   \mathbf{E}_\alpha(r,\phi) = \left[U_\alpha(r,\phi)+U_0(r,\phi)\right]\xhat +i \left[U_\alpha(r,\phi)-U_0(r,\phi)\right].  
 \end{equation}

We may then use Eq.~(\ref{5}) to express the fractional order field $U_\alpha(r,\phi)$ in terms of integer order beams, as in the scalar case.

The topological index of polarization singularities is determined by the change of the phase $\Psi$ of the Stokes vector\cite{gjg:so:2017}  $S_1 + iS_2$ in a closed loop around the singularity, where $\Psi$ is given by
\begin{equation}\label{10}
    \begin{split}
        \Psi =& \frac{1}{2}\tan^{-1}\left(\frac{S_2}{S_1}\right) , \\
        S_1 =& E^2_{x} - E^2_{y}, \\
        S_2 =& 2 Re\left[E_{x}E^*_{y}\right].
    \end{split}
\end{equation}

From Eq.~(\ref{E:pm}) and the discussion of scalar phase singularities, we expect that the total topological index will increase by 1 when the fractional order $\alpha=1$, with two $n=1/2$ polarization singularities created in the transition. 

\subsection{Topological charge and index}

The topological charge $t$ or index $n$ of a singularity can be formally defined by the expression,
\begin{equation}\label{11}
    t,n = \frac{1}{2\pi}\oint _C \nabla \Psi (\mathbf{r})\cdot d\mathbf{r}
 \end{equation}
where $C$ is a closed path of integration around the singularity line and $\Psi(\mathbf{r})$ is the phase of the scalar field or the Stokes vector (orientation angle) for phase and polarization singularities, respectively.

For a scalar field, the topological charge $t$ of an output field corresponding to a fractional singularity $\alpha$ can be written in the simple form,
\begin{equation}\label{12}
    t = m\times\mbox{floor}\left[\frac{\alpha}{m} + \frac{1}{2}\right],
\end{equation}
where ``$\mbox{floor}(x)$'' refers to the largest integer not exceeding $x$.  Equation (\ref{12}) implies that jumps in topological charge of a scalar field can be achieved by modulating the fractional order $\alpha$ of the SPP exhibiting Hilbert's Hotel evolution of phase and polarization singularities.  From Eq.~(\ref{E:pm}), we may conclude that every new scalar vortex in $U_\alpha$ of unit topological charge produces a C-point of $n=1/2$ topological index.  
\begin{figure*}[t]
\centering
\includegraphics[width=0.9\linewidth]{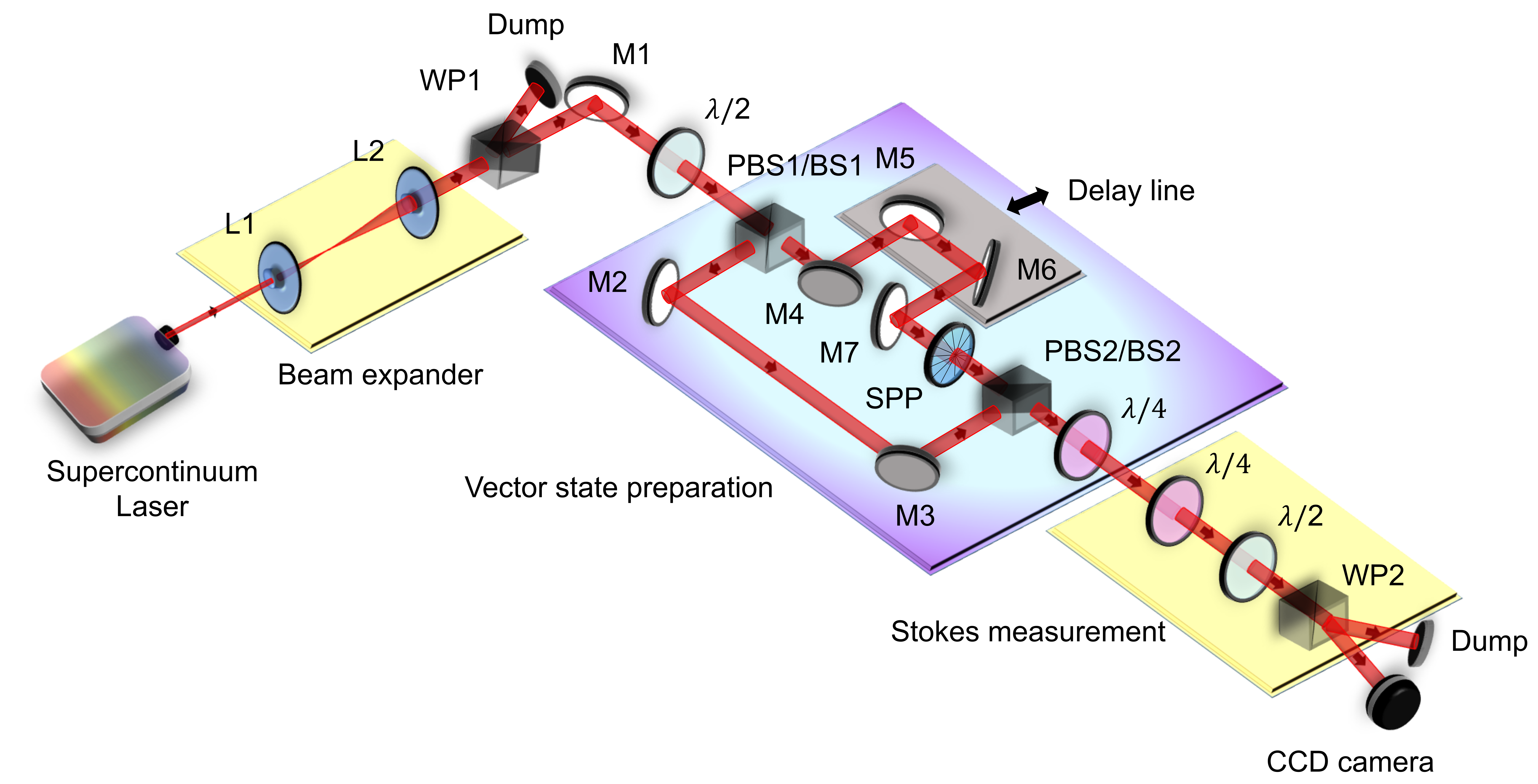}
\caption{ Experimental setup for the observation of optical Hilbert Hotel in phase and polarization. L1-2, Plano-convex lenses; WP1-2, Wollaston prisms; M1-7, dielectric mirrors; $\lambda$/2, half-wave plate; BS1-2, Beam splitter cubes; PBS1-2, Polarising beam splitter cubes; SPP, Spiral phase plate of order 2; $\lambda$/4, quarter-wave plate; CCD, charge-coupled device camera.}
\label{Figure 1}
\end{figure*}
\section{Experimental Setup}

\label{experimentsetup}

The schematic of the experimental setup for the realization of the Hilbert hotel is shown in Fig. \ref{Figure 1}. A supercontinuum laser (NKT Photonics) producing unpolarized radiation with an average power of 2 W tunable across 400 nm to 2200 nm is used as the primary laser for the experiment. Using a variable tunable filter, we tuned the laser wavelength in the visible range across 400 nm to 800 nm with a minimum laser bandwidth of 10 nm. The coherence length of the laser is calculated to be $\sim$ 50 $\mu$m at 700 nm. Using a beam expander comprised of pair of plano-convex lenses L1 and L2 of focal lengths $f1$ = 50 mm and $f2$ = 300 mm, respectively, we have collimated the Gaussian beam ($TEM_{OO}$ mode) of a diameter (full width at half maximum) of $\sim$ 6 mm.  The Wollaston prism (WP1) with a polarization extinction ratio of 100000:1 is used to extract the linear polarized beam corresponding to an average power of 10 mW for a bandwidth of 10 nm across the tuning range. 
A $\lambda/2$-plate is used to control the laser power in the modified polarization Mach–Zehnder interferometer (MZI), comprised of PBS1, PBS2, and a set of plane mirrors, M1-6. We used a delay line in one of the arms in order to temporally overlap the two beams at the output of the MZI. A double-ramp SPP made of BK7 glass having a topological charge of $l$ = 2 at the designed wavelength of 1064 nm was kept in one arm of the MZI to generate a vector vortex beam. 

The electric field at the output of the MZI can be written as 
\begin{equation}\label{13}
\BE_1(r,\phi) = a U_\alpha(r,\phi)\xhat+b U_0(r,\phi)\yhat,
\end{equation}
with $\xhat$ and $\yhat$ representing horizontal and vertical polarization, respectively. However, after propagation through the quarter wave-plate ($\lambda/4$), the electric field of the vector beam transformed into 
\begin{equation}\label{14}
\BE_2(r,\phi) = a U_\alpha(r,\phi)\ehat_++b U_0(r,\phi)\ehat_-,
\end{equation}
where again $\ehat_+$ and $\ehat_-$ are again the unit vectors of left- and right circular polarization of light, and $a$  and $b$ are the relative weights of the two orthogonally-polarized beams. The set of elements $\lambda/4$, $\lambda/2$, and WP2 are used to measure the Stokes parameters of the vector beam. On the other hand, to study the scalar fractional vortex beam, we redesigned the MZI by replacing  the PBS1 and PBS2 with 50:50 beam splitters, BS1 and BS2, respectively. In such a configuration, the arm of the MZI having the SPP generates a fractional vortex beam, while the second arm acts as the reference Gaussian beam or planar wavefront for interference study.

\section{RESULTS AND DISCUSSION}
\begin{figure}[t!]
\centering
\includegraphics[width=0.9\linewidth]{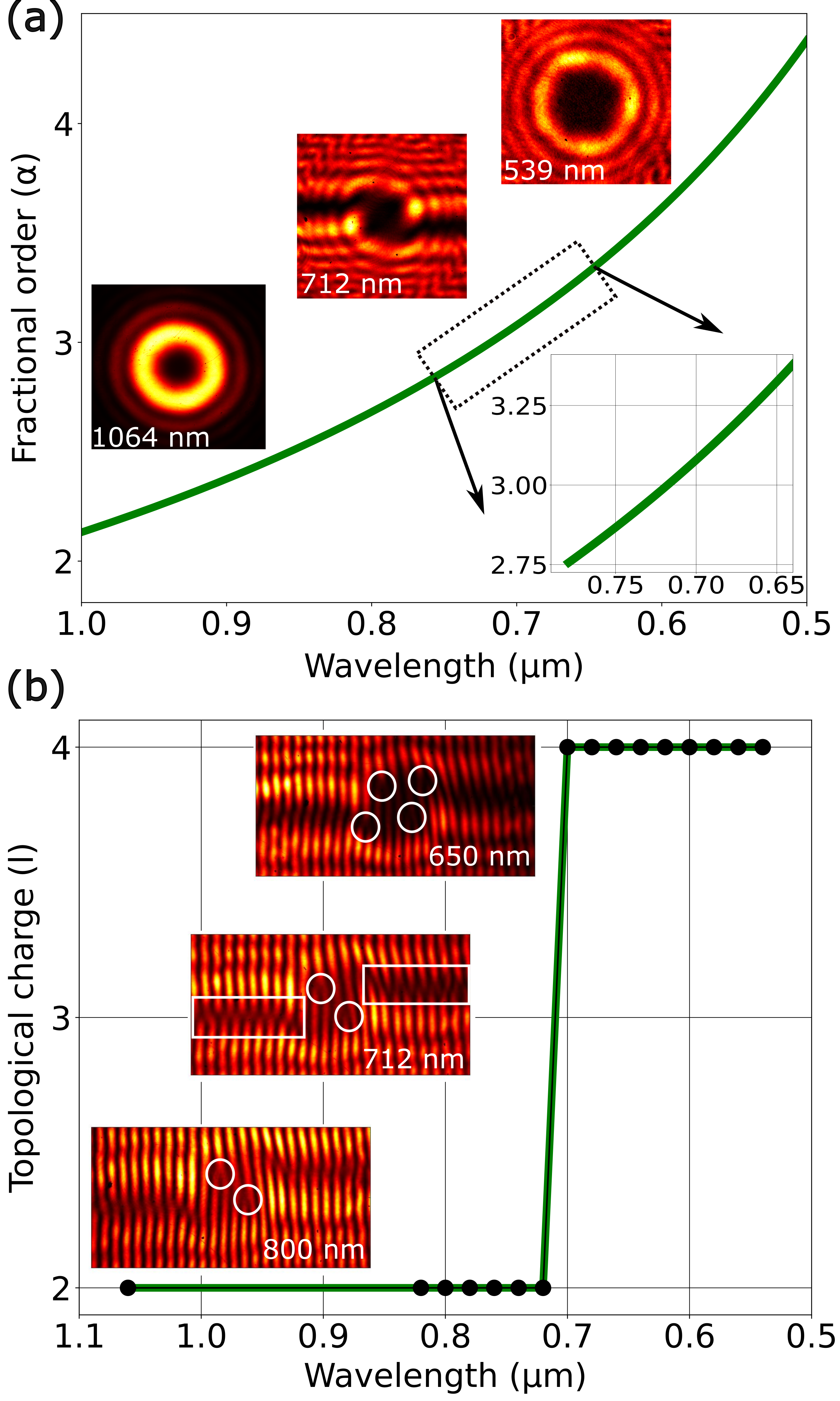}
\caption{(a) Theoretical variation of the fractional order, $\alpha$ of the SPP as a function of the laser wavelength. (Inset) Magnified section of the topological charge variation near 712 nm and the intensity pattern of the beam after SPP at 1064, 712,  and 539 nm. (b) Experimental variation of topological charge, $l$ with wavelength showing step jump near 712 nm. (Inset) Interference pattern showing the characteristic fork pattern of the vortex beam at 800, 712, and 650 nm.}
\label{Figure 2}
\end{figure}

We discuss the scalar case first. Using the dispersion relations of N-BK7 glass \cite{BK7} in Eq.~(\ref{6}) for the double-ramped SPP ($m=2$), we have calculated the effective fractional order $\alpha$ of the SPP as a function of laser wavelength; the results are shown in Fig. \ref{Figure 2}(a). As evident from the figure, the fractional order of the output beam continuously changes from the integer order, $\alpha$ = 2 to $\alpha$ = 4 as the laser wavelength changes from 1064 nm to 539 nm. We have recorded the intensity profile of the output beam at 1064 nm and 539 nm, as shown by the insets of Fig. \ref{Figure 2}(a), which have the clear doughnut intensity profiles of integer vortices. By measuring the characteristic fork intensity interference pattern of the doughnut beam with the reference plane wavefront, we confirm the order of 
the generated vortex beam to be $\alpha$ = 2 and 4 at 1064 nm and 539 nm, respectively. However, according to Eq.~(\ref{12}), a transition of the field topological charge should occur when the plate fractional order $\alpha$ = 3. Therefore, we have magnified the wavelength range from 650 nm to 750 nm as shown by the inset of Fig. \ref{Figure 2}(a), and find the required laser wavelength for $\alpha$ = 3 to be $\lambda$ = 712 nm. Subsequently, we have observed the intensity profile of the beam, which is expected and confirmed to carry two expected singular lines for a $m$ = 2-step SPP. 

To evaluate how the topological charge of the field changes with the variation of  wavelength, we recorded the intensity interference pattern of the output beam with the reference beam and counted the net number of forks while varying the laser wavelength. The results are shown in Fig. \ref{Figure 2}(b). As expected from the theory and Ref. \cite{gg:o:2016}, we observe the net number of forks (marked by white circles) of the interferogram of the vortex beam for the wavelength range from 1064 nm to 720 nm to be two even as the  intensity pattern of the vortex beam changes from doughnut shape to carrying two low-intensity lines. Similarly, the net number of forks of the interference pattern in the wavelength range 700 nm to 539 nm is four despite the change of intensity pattern from carrying low-intensity lines to a doughnut shape. Therefore, we can easily conclude the topological charge of the vortex beam to be $l=2$ for the wavelength range of $1064 - 720$ nm and $l=4$ for the wavelength range of $700 - 530$ nm. 

At the wavelength 712 nm, corresponding to $\alpha = 3$, the interference pattern contains an in principle infinite number of fork pairs representing vortex dipoles lying along the two singular lines of the SPP, as predicted by Berry \cite{mvb:joa} and Gbur \cite{gg:o:2016}. When an infinite number of dipoles are present, the topological charge of the field is indeterminate. However, from Fig. \ref{Figure 2}(b), it is evident that the topological charge of the vortex beam generated by two-step SPP has a two-step jump from $l=2$ to $l + m = 4$ while the laser wavelength drops below 712 nm in agreement with Eq. (\ref{12}). 

These observations suggest the experimental realization of Hilbert's Hotel. They also demonstrate the simplicity and elegance of the overall experiment; it is very easy to select any fractional order between $l=2$ to $l=4$ by adjusting the laser wavelength to the SPP without moving any optical elements. However, due to the material dispersion and the ultrafast nature of the laser, one needs to ensure the temporal overlap of the beams of MZI using the delay stage. Such delay adjustment can be avoided using a continuous-wave laser or laser of a broader pulse width.
\subsection{Scalar fractional vortex beam}
\begin{figure*}[t!]
\centering
\includegraphics[width=\textwidth]{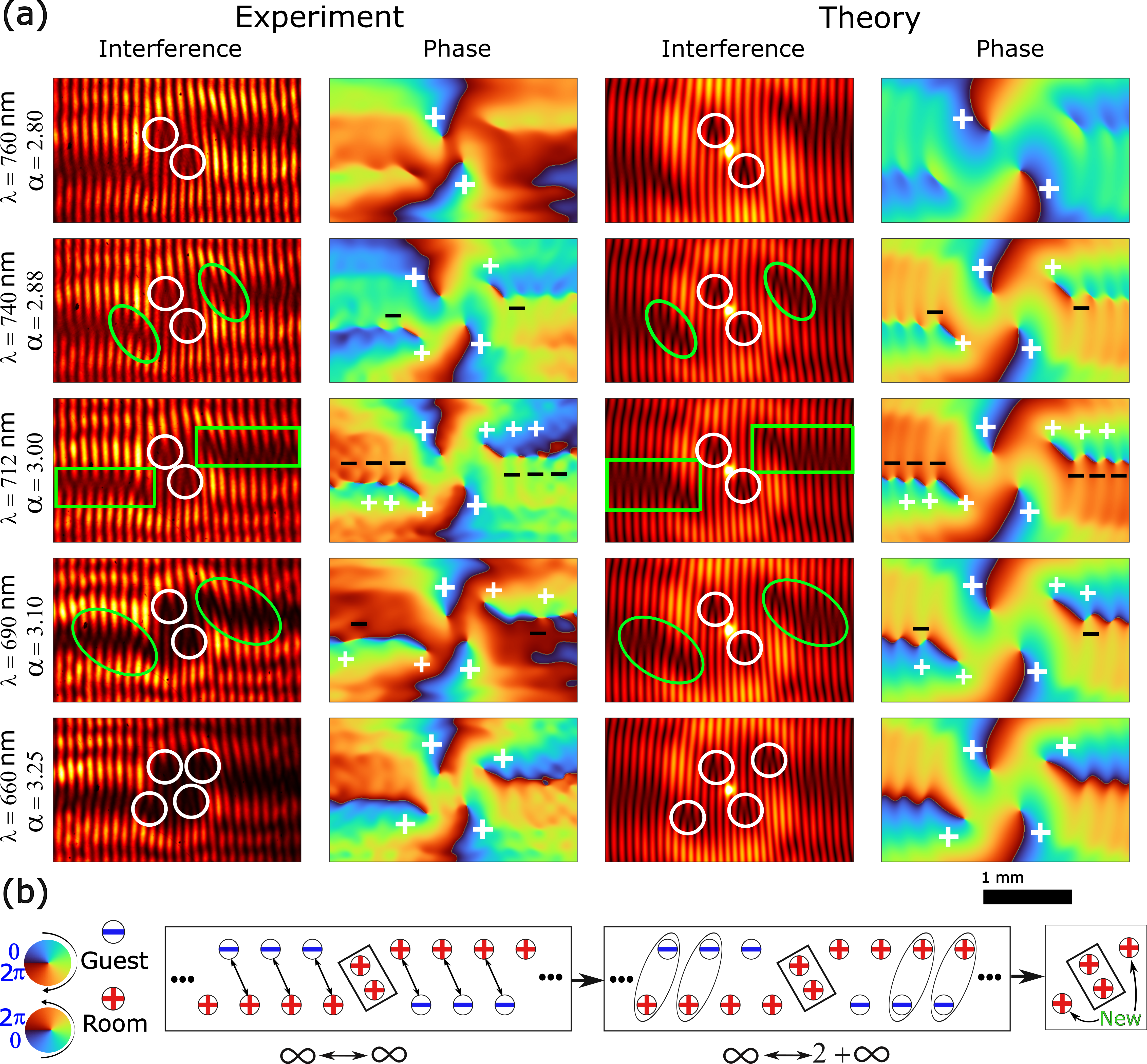}
\caption{(a) Experimental interference pattern (first column) and corresponding phase distribution extracted using Fourier spectrum analysis (second column), theoretical interference pattern (third column), and corresponding phase distribution (fourth column) of the fractional vortex generated for different laser wavelength. The unit fork patterns are marked with white circles, and the dipoles are marked using green ovals or rectangles. The clockwise and counterclockwise phase variations are marked by "-" and "+" signs. (b) Illustration of vortex dynamics with the laser wavelength mimicking Hilbert's Hotel. The phase variation "+" and "-" are labelled by "room" and "guest", respectively. }
\label{Figure 3}
\end{figure*}

To truly demonstrate Hilbert's Hotel, however, we need to show the rearrangement of positive and negative charge vortices in the manner that "rooms" and "guests" are rearranged in Hilbert's thought experiment. Therefore, we have focused our investigation on the wavelength range 760 - 660 nm in the immediate neighbourhood of the wavelength where the fractional order of the field makes the transition from order $\alpha$ = 2 to 4. The results are shown in Fig.~\ref{Figure 3}. We have recorded the interference pattern of the vortex beam for five different wavelengths, $\lambda$ = 760, 740, 712, 690, and 660 nm, corresponding to the topological orders calculated from Fig. \ref{Figure 2}, of $\alpha$ = 2.8, 2.88, 3, 3.1, and 3.25 respectively, with the results shown first column of Fig. \ref{Figure 3}. 

We observe two forks for $\lambda$ = 760 nm; however, the decrease in laser wavelength to 740 nm results in the creation of fork pairs (upward and downward-opening fork-shaped fringes) corresponding to vortex dipoles (marked by the green ellipses) in the singular lines of low intensity. The series of fork pairs (marked by the green boxes) extends in principle to infinity (here, restricted to the spatial extent of the beams) for a laser wavelength of 712 nm. A further decrease in laser wavelength results in the annihilation of the vortex dipoles,
as evident from the reduction in the number of fork pairs at 690 nm. Finally, it attains a net of four forks, corresponding to the topological charge of 4 at 660 nm. 

To understand and appreciate the creation and annihilation of the vortex dipole while changing the laser wavelength to the SPP, we have extracted the phase of the beam from the interference pattern by using Fourier spectrum analysis \cite{Takeda:82} of a non-contour type fringe pattern. The results are shown in the second column of Fig.~\ref{Figure 3}. 

It is evident that the phase distribution of the vortex beam at 760 nm has two phase singularities marked by "+" at the center of the beam with phase varying from 0 to 2$\pi$ in the anticlockwise direction. The decrease in wavelength to 740 nm results in the creation of a pair of vortices in the singular lines marked by "+" and "-" (phase varying from 0 to 2$\pi$ in the clockwise direction) on either side of the existing two vortices at the beam center. However, at 712 nm, corresponding to the topological order of $\alpha$ = 3, we see an increase in the creation of vortex pair extending away from the center of the beam. Due to the finite size of the beam and the apertures of the CCD camera, we have restricted our study to three pairs on either side of the beam center. In principle, however, the number of pairs extends indefinitely, eventually becoming unmeasurable in the low-intensity outskirts of the beam.

As predicted in Ref. \cite{gg:o:2016}, we experimentally observe each vortex to annihilate with its opposite neighbour instead of the neighbour it was created with. We see this occur starting at the laser wavelength of 690 nm, corresponding to fractional order $\alpha$ = 3.10, and progressing towards the origin from the most distant points with further decrease of laser wavelength (an increase of topological order away from $\alpha$ = 3), leaving four "+" vortices at the center of the beam at a wavelength of 660 nm. Using the experimental parameters in Eq.~\ref{5}, we have simulated the interference fringes and extracted the phase distribution with the results shown in the third and fourth columns of Fig. \ref{Figure 3}, finding them in excellent agreement with the experimental results. 

To make a clear relationship between the experimental results and Hilbert's Hotel, we have labelled the phase singularities marked by "+" and "-" as the "room" and "guest" of the Hotel and illustrated the vortex dynamics using the phase distribution of the beam with fractional order $\alpha$ = 3, 3.10 and 3.25. The results are shown in Fig. \ref{Figure 3}(b). As evident from Fig. \ref{Figure 3}(b), the phase distribution has two individual charges at the center and a series of vortex pairs in the singular lines on either side of the beam center for $\alpha$ = 3. Those vortex pairs created together are connected by arrows. Let us ignore the unit charges (contained in the black box) present at the center of the beam and focus on the vortex dipoles that mimic the situation of a fully occupied Hilbert Hotel, i.e., $\infty \leftrightarrow$  $\infty$. Due to the experimental constraint, we have restricted our study to three pairs of dipoles. However, for $\alpha$ = 3.10, we observe the annihilation of the vortex dipole pairs of the series away from the beam center, leaving the first vortex dipole and a unit charge "+" from the second pair. The existence of unit charge "+" of the second pair and the increase of the separation of vortex charge "+" and "-" of the first dipole confirm that the vortex charge "-" of each dipole has annihilated with the vortex charge "+" of the next neighbouring dipole, i.e., the vortex charge "-" of N$^{th}$ dipole annihilates with vortex charge "+" of the (N+1)$^{th}$ dipole, as if the "guest" of one "room" has moved to the adjacent "room". Finally, for $\alpha = 3.25$, we see that the vortex charge "-" ("guest") of the first dipole has annihilated with vortex charge "+" ("room") of the second dipole, leaving an extra vortex charge of "+" ("room") on either side of the beam mimicking the creation of two vacant "rooms" for the new "guests", $\infty \leftrightarrow$  2 +  $\infty$, in the fully occupied Hilbert Hotel. This is, to the best of our knowledge, the first demonstration of an optical Hilbert's Hotel for a scalar field using a multi-ramp SPP.  It demonstrates that a multi-ramp SPP can in fact produce multiple new vortices simultaneously, just as it is possible to open up multiple rooms in Hilbert's Hotel by asking each guest to move more than one room over.
\begin{figure}[t!]
\centering
\includegraphics[width=\linewidth]{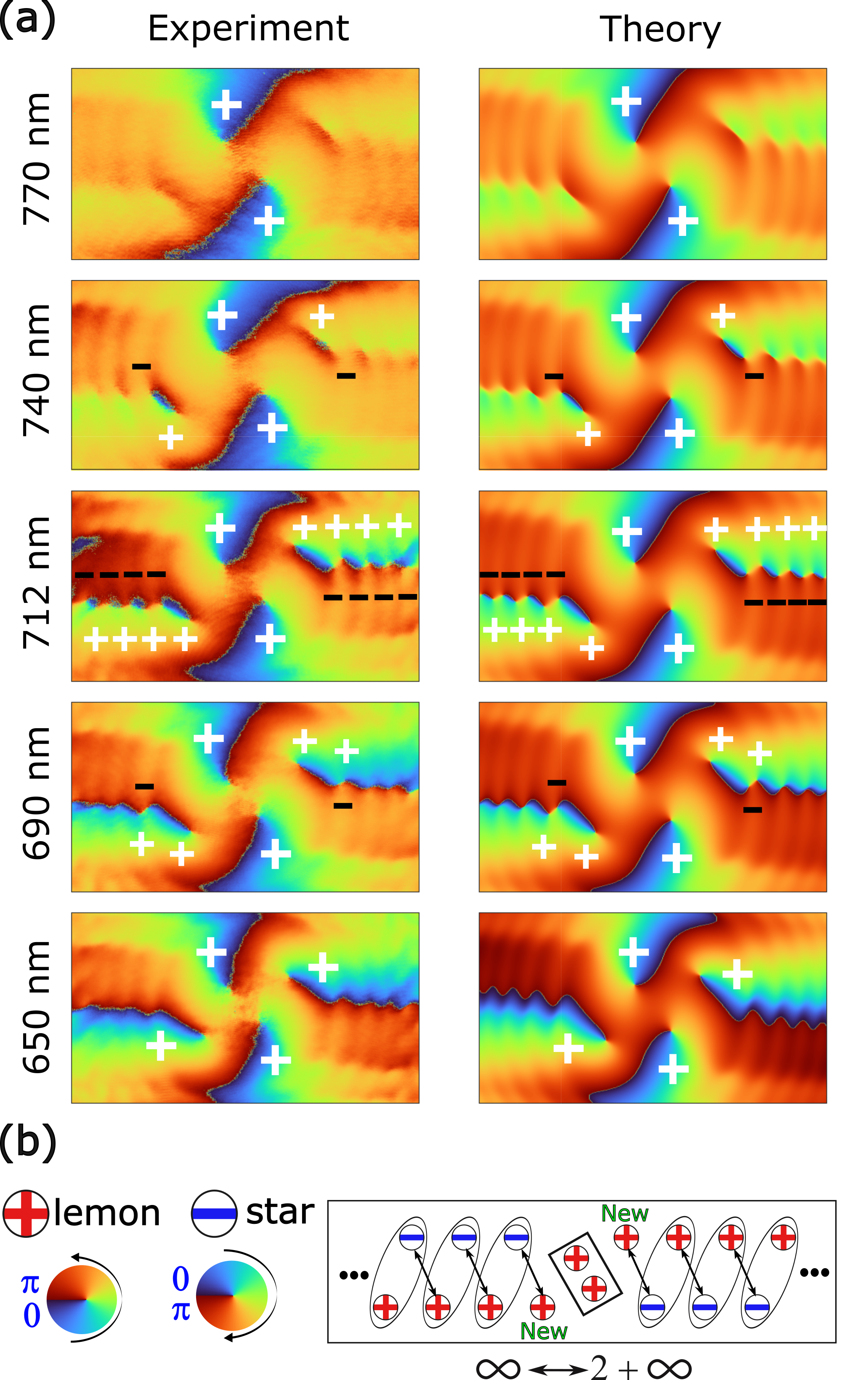}
\caption{(a) Experimental (first column) and theoretical (second column) dynamics of polarization singularity of fractional vector beam with laser wavelength. The polarization ellipse orientation (0 - $\pi$) about the singular point in the counterclockwise and clockwise directions are identified by "+" and "-" respectively. (b) Illustration of dynamics of polarization singularity, "+" (lemon) and "-" (star) with the laser wavelength mimicking $\infty \leftrightarrow$  $\infty$, to $\infty \leftrightarrow$  2 +  $\infty$, transition of Hilbert Hotel.}
\label{Figure 4}
\end{figure}

\subsection{Vector fractional vortex beam}
We now turn to the case of Hilbert's Hotel with polarization singularities in a vector beam. It is known that the coaxial superposition of a vortex and a Gaussian beam, with opposite circular polarizations, transforms the phase singularity of a scalar vortex beam into a polarization singularity \cite{beckley2010full} of a vector vortex beam. Here the polarization singularities, or C-points, are the points on the vector beam where the polarization ellipse is circular, and the orientation of the polarization ellipse is undefined. The "star" and "lemon" singularities are popular C-point singularities where the polarization ellipse around the singularity point orientations from $0$ - $\pi$ in the clockwise and counterclockwise directions, respectively. We have observed the vector beam to carry the "star" and "lemon" singularities in pairs and to carry topological charges of $+1/2$ and $-1/2$, respectively \cite{Kumar:22}. As a result, in 2017, Wang and Gbur \cite{Wang:17} theoretically demonstrated the Hilbert Hotel using polarization singularities in vector beams. 

Building upon the simple experimental realization of Hilbert Hotel using fractional scalar vortex beams, we have explored the experimental realization of Hilbert Hotel using polarization singularities. In the current experiment (see Fig. \ref{Figure 1}), the replacement of BS1 and BS2 with PBS1 and PBS2 of the MZI results in the vector beam with an electric field given by Eq. \ref{13}. Keeping a $\lambda/4$ plate at +45$^{\circ}$ with respect to the horizontal polarization, we have converted the vector vortex beam from a linear basis to a circular basis with a generalized form represented by Eq. (\ref{14}). Similar to the scalar vortex study (see Fig. \ref{Figure 3}, we have adjusted the laser wavelength at 770, 740, 712, 690, and 650 nm and recorded the beam intensity profile for four distinct configurations of $\lambda/4$ and $\lambda/2$ plates, adequate for estimating the Stokes parameters of the polarisation distribution \cite{goldstein2017polarized}. 

Using these intensity distributions, we have calculated the ellipse orientation distribution with the help of Stoke's parameters, S1 and S2, given by Eq. (\ref{10}). The results are shown in Fig. \ref{Figure 4}. As evident from the first column of Fig. \ref{Figure 4}(a), the polarization ellipse orientation map of fractional vector vortex in circular basis has two lemon singularities denoted by "+" at the center of the beam for laser wavelength $\lambda$ = 770 nm. However, for laser wavelength of $\lambda$ = 740 nm, we observe, in addition to the initial lemons, the appearance of pairs of lemon and star singularities identified by "+" and "-", respectively, on either side of the beam center. The number of lemons and stars keeps on increasing with the decrease of laser wavelength, in principle giving a countably infinite number of pairs extended away from the beam center for the laser wavelength of $\lambda$ = 712 nm corresponding to the topological order $\alpha$ = 3 of the fractional vector beam of the two-step SPP. 

However, as we further reduce the laser wavelength to 690 nm, we observe, similar to the vortex pair annihilation of scalar fractional case as shown in Fig. \ref{Figure 3}, the annihilation of star and lemon singularities; star of N$^{th}$ pair annihilates with the lemon of (N+1)$^{th}$ pair, leaving one pair of lemon and star and a single lemon singularity of the adjacent pair on either side of the beam center. Finally, the star of the first pair annihilates with the lemon of the second pair creating an extra lemon singularity on either side of the beam center with a result of a total of four lemons, including the initial lemons for the laser wavelength of $\lambda$ = 650 nm corresponding to $\alpha$ = 3.25. 

Using the mathematical Eqs.~(\ref{E:pm}), (\ref{10}) and (\ref{14}) and the experimental parameters, we have calculated the evolution of polarization singularity of the fractional vector beam as shown in the second column of Fig. \ref{Figure 4}(a) in close agreement with the experimental results. The evaluation of polarization singularities as summarized in Fig. \ref{Figure 4}(b) show the transition of  $\infty \leftrightarrow$  $\infty$, to $\infty \leftrightarrow$  2 +  $\infty$, the transition of Hilbert Hotel. Therefore, we have successfully observed the first experimental realization of more general examples of Hilbert’s Hotel with polarization singularities in vector beams, as predicted theoretically \cite{Wang:17}.

\begin{figure}[htbp]
\centering
\includegraphics[width=\linewidth]{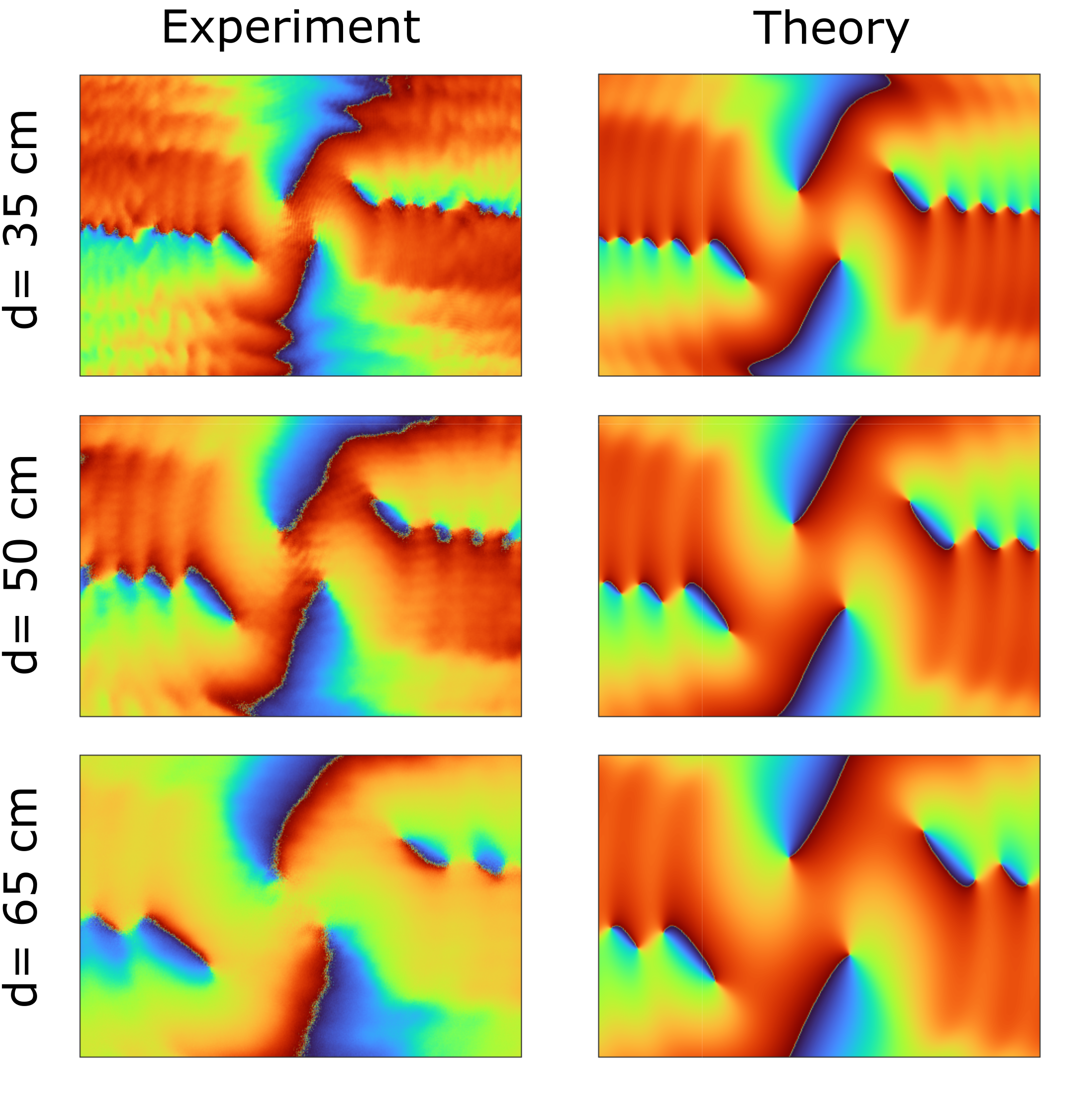}
\caption{Experimental (first column) and theoretical (second column) evolution of the polarization singularities chain of the fractional vector beam corresponding to a topological order of $\alpha$ = 3 at 712 nm at three different propagation distances in free space. 
}
\label{Figure 5}
\end{figure}

We have further studied the propagation characteristics of the polarization singularity of the fractional vector vortex. In doing so, we measured the beam intensity for different projections at different distances along the beam propagation for the laser wavelength of $\lambda$ = 712 nm ($\alpha$ = 3) and extracted the ellipse orientation distribution. The results are shown in Fig. \ref{Figure 5}. As evident from the first column of Fig. \ref{Figure 5}, the ellipse orientation map contains the infinitely extended chain of polarization singularities at the propagation of 35 cm measured from the MZI. On further propagation to a distance of 50 cm and 65 cm, we observe the positions and size of singularity change significantly due to the beam divergence. However, the infinite polarization singularity chains remain unchanged and preserve the signature of Hilbert's Hotel. It is to be noted that, due to the divergence of the finite beam, the infinite line of phase singularities at a long propagation distance is lost in the low-intensity region of the beam tail. Using the same experimental parameters, we have also simulated the propagation characteristics of the infinite lines of polarization singularity chains as shown in the second column of Fig. \ref{Figure 5} in close agreement with the experimental results. 

\section{Conclusions}
In this paper we have demonstrated, to the best of our knowledge, the first experimental realization of Hilbert's Hotel using both fractional order scalar and vector beams with a multi-ramp SPP. The generation of fractional scalar and vector vortex beams from a fixed SPP by changing the laser wavelength through a filter makes the overall experiment very simple, alignment-free, and easy to implement. While we have demonstrated the proof of concept using a SPP of order $l$ = 2,  the current demonstration reveals that the use of multi-ramp SPP can produce multiple new vortices simultaneously. The generation of such multiple vortices can open the possibility of verifying the complicated transitions to prove the generalized Hilbert's Hotel paradox. The generic experimental scheme can also be useful for understanding the behaviour of complex polarization-sensitive optical elements as required in many fields, including designing novel devices, quantum communication, and sensing.
\section*{AUTHOR DECLARATIONS}
\subsection*{Conflict of Interest}
The authors have no conflicts to disclose.
\subsection*{Author Contributions}
S. K., A. G., and C. K. developed the experimental setup and performed measurements, data analysis, and numerical simulation. A. G. lead the experiment and wrote the original draft. A. S. and G. G. lead the theoretical study, derived analytical formulas, and numerical simulation. S. S. managed the laser and participated in the discussion. G. S. developed the ideas and lead the project. All authors participated in the discussion and contributed to the manuscript writing.

\begin{acknowledgments}
We wish to acknowledge the support of Mr. B. S. Bharath Saiguhan, Physical Research Laboratory, Ahmedabad, in the coding.
\end{acknowledgments}
\section*{DATA AVAILABILITY}
The data that support the findings of this study are available from the corresponding author upon reasonable request.

\section*{REFERENCES}
\bibliography{references}
\end{document}